\documentclass[a4paper,12pt,aps,superscriptaddress]{article}
\usepackage{amsmath}

\usepackage{amsfonts}
\usepackage{amssymb}

\usepackage{float}
\usepackage{epsfig}

\newcommand{\be}{\begin{equation}}
\newcommand{\ee}{\end{equation}}
\newcommand{\bea}{\begin{eqnarray}}
\newcommand{\eea}{\end{eqnarray}}

\begin{document}
\title{\quad\quad\quad	Statistical Curse of the  Second Half Rank}
\author{Jean Desbois$^{1,2}$, St\'ephane Ouvry$^{1,2}$, Alexios Polychronakos$^{3}$\\
$^{1}$ Universit\'e  Paris-Sud, Laboratoire de Physique Th\'eorique et Mod\`eles
Statistiques\\  
  UMR 8626, F-91405 Orsay  \\
$^{2}$ CNRS, LPTMS,  F-91405 Orsay \\
$^{3}$ Physics  Department, City College of New York, NY\\}
\maketitle

\begin{abstract}
In competitions involving many participants running many races the final rank is determined by the
score of each participant, obtained by
adding its ranks in each individual race. The ``Statistical Curse of the  Second Half Rank" is the observation that if the score of a participant is even modestly worse than
the middle score, then its final rank will be much worse (that is, much  further away from the middle rank) than might have been expected. We give  an explanation of
this effect for the case of a large number of races using the Central Limit Theorem. We present exact quantitative results in this limit and demonstrate that the score probability distribution will be gaussian with scores packing near the center. We also
derive the final rank probability distribution for the case of two races and we present some exact
formulae verified by numerical simulations for the case of three races. The variant in which the
worst result of each boat is dropped from its final score is  also analyzed and solved for the
case of two races.     

\medskip\noindent {}

\end{abstract}

\vskip 0.2cm

{\bf }

\vskip 0.2cm

\section{Introduction} 

In competitive individual sports involving many participants it is { in some cases} standard practice
to have several races and determine the final rank for each participant by taking  the sum of its ranks in each
individual race,    thereby defining  its score. By comparing the scores of the participants  a final rank  can be decided among them. Typical examples are  regattas, which can
involve a large number of sailing boats ($\sim$100), running a somehow large number of consecutive
races ($\ge 10$).

An empirical observation of long-time participants is that, if their scores are even slightly
below the average, their final rank will be much worse than expected. This frustrating fact,
which we may call the ``Statistical Curse of the Second Half Rank", is analyzed in this
work and argued to be due to statistical fluctuations in the results of the races, on
top of the inherent worth of the participants.  Using some simplifying assumptions we
demonstrate that it can be explained by a version of the Central Limit Theorem \cite{1,2}  for correlated random variables. A general result for a large number
of participants and races is derived. Some exact
resuts for a small number of races are presented. A variant of the problem, in which the
worst score for each participant is dropped, is also considered and solved for the case of two races.

\section{Basic setup}
 
Consider $n_b$ boats racing $n_r$  races. A boat $i$ in the  race $k$ has an individual rank $n_{i,k}\in [1,n_b]$ (lower ranks represent better performance). The  score  of the boat $i$ is the sum $n_i=\sum_{k=1}^{n_r} n_{i,k}\in [n_r,n_r n_b]$ of its individual ranks in each  race.  The  final rank of boat $i$  is determined by the place occupied by its score $n_i$ among the scores
of the other boats $n_j$, with $j\ne i$.

For reasons of simplicity we assume that in a given race the ranks are uniformly distributed random  variables with  no exaequo (that is, all boats are inherently equally worthy and there are no ties).
We shall also take the ranks in different races to be  independent random variables.  It follows that for the race $k$ the set $\{n_{ik}; i=1, 2,..., n_b\}$ is a random permutation of $\{1, 2,..., n_b\}$  so that the $n_{i,k}$'s are correlated random variables (in  particular  $\sum_{i=1}^{n_b}n_{i,k}=n_b(n_b+1)/2$), while $n_{ik}$ and
$n_{jk'}$ are uncorrelated for $k \neq k'$. We are interested in the probability distribution 
for boat $i$ to have a final rank $m\in [1,n_b]$ given  its score. 

Let us illustrate this situation in the simple case of three boats racing two races. We have to take all random permutations of $\{1,2,3\}$  both for the first and the second race, and to add them  to determine   the possible scores of the three boats.  It is easy to see that for, say,  boat  $1$ to have  a score $n_{1,1}+n_{1,2}=4$ there are twelve possibilities:

\noindent
 i)  four instances   where  $n_{1,1}=1$ and $n_{1,2}=3$,

\noindent
ii) four instances where  $n_{1,1}=2$  and $n_{1,2}=2$, and

\noindent
iii) four instances where $n_{1,1}=3$  and $n_{1,2}=1$.

\noindent
In each of these three cases (i), (ii) and (iii), one finds that boat $1$ has an equal probability $1/2$ for    
its final rank to be either $m=1$ or $m=2$. Its mean rank  follows as $\langle m\rangle =1/2(1+2)=3/2$.  Clearly the  score $4$ is precisely  the middle  of the set $\{2,3,4,5,6\}$ and  $\langle m\rangle =3/2$ is indeed close\footnote{The $1/2$  discrepancy is due to the fact that
boats with equal scores are all assigned the same final rank. E.g., two boats tying in the
first place are assigned a rank of 1, while the next boat would have a rank of 3. If, instead, the two top boats were assigned a rank of 1.5 (the average of 1 and 2) we would have obtained
$\langle m\rangle =2$. This effect, at any rate, will be important only for a small number
of boats. } to the middle rank $2$. 

More interestingly, cases (i), (ii) and (iii) give the same final rank probability distribution. It means that the final rank probability distribution depends only on the score of  boat $1$, and not on its individual ranks in each of the two races consistent with its  score. This fact is particular to two races and would not be true any more for three or more races. The final rank probability distribution for  boat $1$ given its score would depend in this case
on the full set of its ranks in each race, and not just on its score. The final rank probability distribution should then be defined as the average of the above distributions for all set of ranks consistent with its  score.

To  avoid this additional averaging and simplify slightly the analysis, we consider from now on  $n_b$ boats racing $n_r$  races, plus an additional  virtual boat which is only specified by  its score $n_{t}\in[n_r,n_r( n_b+1)]$. We are interested in finding the probability distribution for this  virtual boat to have  a final rank $m\in [1,n_b+1]$ given  its score $n_{t}$ when it is compared to the set of scores $\{n_i; i=1,2,...,n_b\}$ of the $n_b$ boats. By definition this probability distribution will then depend only on three variables: $n_b$, the number of boats; $n_r$, the number of races; and $n_t$, the score of the virtual boat we are interested in.

\section{The limit of many races}

The problem simplifies when some of the parameters determining the size of the system become large so that we can use central limit-type results. 
In this section we consider the limit in which the number of races becomes large.

We start with a reminder of the Central Limit Theorem in the case of correlated random variables. 
Assume $\{x_{i,k}; i=1,\dots,n_b; k=1,2,\dots,n_r\}$ to be correlated random variables such that
\begin{itemize} 
\item they are independent for different $k$,   
\item the set $\{x_{1,k},x_{2,k},...,x_{n_b,k}\}$ is distributed according to a joint density probablility distribution which is $k$-independent and whose  first two moments (mean and covariance)   are $\langle x_{i,k}\rangle=\rho_i$ and $\langle x_{i,k}x_{j,k}\rangle-\langle x_{i,k}\rangle\langle x_{j,k}\rangle=\rho_{ij}$. 
\end{itemize}

The CLT states that  in the limit $n_r \gg 1$ the summed variables  $x_i=\sum_{k=1}^{n_r} x_{i,k}$  are  correlated gaussian random variables with
$\langle x_i\rangle=n_r \rho_i$ and $\langle x_{i}x_{j}\rangle-\langle x_{i}\rangle\langle x_{j}\rangle=n_r\rho_{ij}$, that is, they  are distributed in this limit according to the probability density
\be
f(x_{1},x_{2},...,x_{n_b})=N\exp[-{1\over 2 n_r}\sum_{i,j}\lambda_{ij}(x_i-n_r\rho_i)(x_j-n_r\rho_j)]
\ee
where $N$ is a normalization constant. The  matrix $[\lambda]$ is the inverse of the covariance matrix $[\rho]$, assuming that $[\rho]$ is non-singular.

In  the race problem,  $x_{i,k}=n_{i,k}$ and $x_i=n_{i}$: one has 

\be\rho_i={n_b+1\over 2}\ee

\be \rho_{ii}={n_b^2-1\over 12}
~,~~~ \rho_{ij}=-{n_b+1\over 12} ~~(i \neq j)
\ee
 (off diagonal correlations are negative) so that  
 
\be \rho_{ij}={n_b+1\over 12}(n_b\delta_{i,j}-1)\quad\quad  i,j\in[1,...,n_b]\ee 

\noindent It follows that in the large  number of races limit $\langle n_i\rangle= n_r{n_b+1\over 2}$  and 
$\langle n_in_j\rangle-\langle n_i\rangle\langle n_j\rangle=n_r\rho_{ij}$.

The covariance matrix $[\rho]$ is singular with a single zero-eigenvalue eigenvector  $(1,1,...,1)$. Any vector perpendicular to   $(1,1,...,1)$, that is, such that the sum of its entries is $0$, is  an eigenvector with eigenvalue $n_r(n_b+1)/ 2$. The fact that $(1,1,...,1)$ is a 
zero-eigenvalue eigenvector signals that the variable  $\sum_{i=1}^{n_b}n_i=n_r n_b(n_b+1)/2$  is deterministic. It must be ``taken out" of the set of the scores before finding the large $n_r$ limit.  We arrive at the density probability  distribution 
\be 
f(n_1,\dots,n_{n_b})=\sqrt{2\pi n_b\over \lambda} \left(\sqrt{\lambda\over 2\pi}\right)^{n_b}\delta\left(\sum_{i=1}^{n_b} n_i-{6\over \lambda}\right)\exp\left[
-{\lambda\over 2}\sum_{i=1}^{n_b}(n_i-n_r{n_b+1\over 2})^2\right]
\ee 
with 
\be\lambda={12\over n_rn_b(n_b+1)}\ee 
 such that indeed $\langle n_i\rangle= n_r\rho_i$  and 
$\langle n_in_j\rangle-\langle n_i\rangle\langle n_j\rangle=n_r\rho_{ij}$.

One can exponentiate the  constraint $\delta(\sum_{i=1}^{n_b}(n_i-n_r (n_b+1)/2)$ so that
 \be 
f(n_1,...,n_{n_b})= \sqrt{n_b \lambda^{n_b -1}\over (2\pi)^{n_b +1}} \int_{-\infty}^{\infty}\exp\left[-ik\sum_{i=1}^{n_b}(n_i-n_r {n_b+1\over 2})-{\lambda\over 2}\sum_{i=1}^{n_b}(n_i-n_r{n_b+1\over 2})^2\right]dk
\ee

For a virtual boat with  score $n_{t}$ the probability  to have a final rank $m$   is
the probability for $m-1$ boats among the $n_b$'s to have a  score $n_i<n_{t}$  and for the other $n_b-m+1$'s to have a score $n_i\ge n_{t}$
 \be 
P_{n_{t}}(m)={n_b\choose m-1}\int_{-\infty}^{n_{t}}dn_1 \dots dn_{m-1}
\int_{n_{t}}^{\infty}dn_{m} \dots dn_{n_b}f(n_1,\dots ,n_{n_b})
\ee 
which obviously satisfies $\sum_{m=1}^{n_b+1} P_{n_{t}}(m)=1$. It can be rewritten as
\be 
P_{n_{t}}(m)={n_b\choose m-1}\int_{-\infty}^\infty w_{n_{t}}(k)^{m-1} (1- w_{n_{t}}(k))^{n_b-m+1}\sqrt{n_b\over 2\pi\lambda}\exp\left[-{n_bk^2\over 2\lambda}\right]dk\label{clt}
\ee 
where 
\be
 w_{n_{t}}(k)=\sqrt{ \lambda\over 2\pi}\int_{-\infty}^{n_{t}}\exp\left[-{\lambda\over 2}(n-n_r{n_b+1\over 2}+{ik\over\lambda})^2\right]dn
\ee
If we further define  
\be
\bar{n}_{t}=\sqrt{\lambda}\left( n_{t}-\frac{n_r(n_b+1)}{2} \right)
\ee
and absorb $1/\sqrt{\lambda}$ in $k$, (\ref{clt}) becomes 
\be
P_{n_{t}}(m)={n_b\choose m-1}\int_{-\infty}^{\infty} {w}_{\bar{n}_{t}}(k)^{m-1}(1- {w}_{\bar{n}_{t}}(k))^{n_b-m+1}\sqrt{n_b\over 2\pi }\exp\left[-{n_b{k}^2\over 2}\right]d{k}\label{cltbis}
\ee
 with
 \be {w}_{\bar{n}_{t}}(k)=\sqrt{1\over 2\pi}\int_{-\infty}^{\bar{n}_{t}}\exp\left[
-\frac{(n+ik)^2}{2}\right]dn\label{w}
\ee

The probability distribution (\ref{cltbis})  is  of binomial form but with a $k$-dependent
`pseudo-probability' $w_{\bar{n}_{t}}(k)$, and $k$ normally distributed according to
$\sqrt{n_b/( 2\pi)}\exp[-{n_bk^2/ 2}]$.
We find in particular
\be
\langle m\rangle = 1+n_b\int_{-\infty}^{\infty} {w}_{\bar{n}_{t}}(k)\sqrt{n_b\over 2\pi }\exp\left[-{n_b{k}^2\over 2}\right]d{k}
= 1 + n_b \, {\cal{N}} \left( \bar{n}_t \sqrt{\frac{n_b}{n_b -1}}\right)
\label{average}
\ee
where ${\cal{N}}(x)$ is the cumulative probability distribution of a normal
variable
\be
{\cal{N}}(x) = \frac{1}{\sqrt{2\pi}} \int_{-\infty}^x \exp\left[ -\frac{n^2}{2}
\right] dn
\ee

\begin{figure}[htbp]
\epsfxsize=8cm
\centerline{\epsfbox{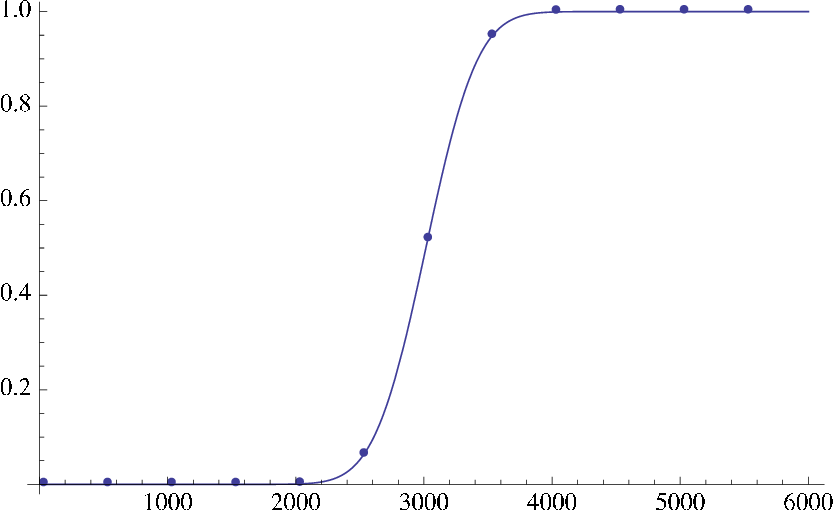}}
\caption{The  final rank of the virtual boat for $200$ boats racing  $30$ races : the continuous line is (\ref{rank}) and the points are numerical simuations for a score $n_t$ ranging from $30$ to $6000$ by steps of $500$. Both data in the curve and the simulation points have been divided by $200$. The "statistical curse (blessing) of the  second (first) half rank" effect is clearly visible on the figure.  }
\label{averagerank}
\end{figure}

We can go further  by considering (\ref{cltbis}) in the large boat number  limit $n_b\gg 1$. In this limit, $n_t$ scales like $n_b$ and thus $\bar{n}_{t}$ is $n_b$-independent: the $n_b$ dependence of $P_{n_{t}}(m)$
is solely contained in the binomial coefficient  and the exponents,  not in ${w}_{\bar{n}_{t}}(k)$. 
Setting $r=m/n_b$ (the percentage rank) and using $n!\simeq \sqrt{2\pi n}(n/e)^n$ we obtain
\be 
P_{n_{t}}(r)={1\over \sqrt{2\pi r(1-r)}}\int_{-\infty}^{\infty} \exp\left[-n_b \left(r\ln{r\over {w}_{\bar{n}_{t}}(k)}+(1-r)\ln{1-r\over 1-{w}_{\bar{n}_{t}}(k)}+k^2/2\right)\right]d{k}\label{cltter}
\ee
In (\ref{cltter}) the exponent of the integrand    
is  negative except  when $k=0$  and $r={w}_{\bar{n}_{t}}(k)$:
for large $n_b$  a saddle point approximation  yields that  $P_{n_{t}}(r)$  vanishes except when $r$ is taken to be ${w}_{\bar{n}_{t}}(0)$.  It follows that the final rank of the virtual boat is essentially fixed by its score $\bar n_t$
\be 
{\bar r} = {\cal{N}}({\bar n}_t )
\label{rank}
\ee
as expected from (\ref{w}, \ref{average}) in the large $n_b$ limit
and shown in Fig. \ref{averagerank} for $200$ boats racing $30$ races.

\begin{figure}[htbp]
\epsfxsize=8cm
\centerline{\epsfbox{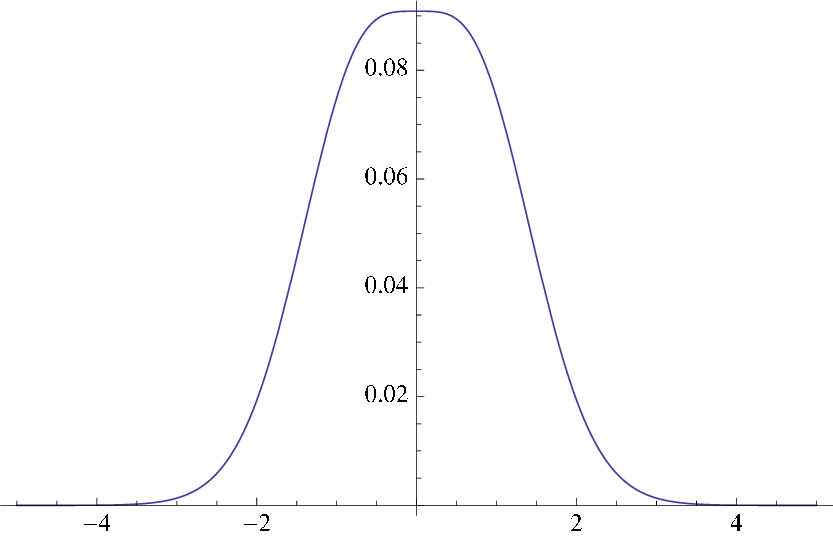}}
\caption{The Kollines function}
\label{Kollines}
\end{figure}

The fluctuations of $r$ around $\bar r$ are obtained by expanding  the exponent in (\ref{cltter})  around $r=\bar r$ (one sets $r\simeq  \bar r+\epsilon$)   and around $k=0  $ 
so that 
\be r\ln{r\over {w}_{\bar{n}_{t}}(k)}+(1-r)\ln{1-r\over 1-{w}_{\bar{n}_{t}}(k)}\simeq {(\epsilon-k {w}'_{\bar{n}_{t}}(0))^2\over 2\bar r(1-\bar r)}\ee
where ${w}'_{\bar{n}_{t}}(0)$ is the derivative of ${w}_{\bar{n}_{t}}(k)$ at $k=0$.
The integration over $k$ in (\ref{cltter})  finally yields
\be 
P_{n_{t}}(r)={1\over\sqrt{2\pi n_b(\bar r(1-\bar r)+{w}'_{\bar{n}_{t}}(0)^2)}}\exp\left[-{n_b\epsilon^2\over 2(\bar r(1-\bar r)+{w}'_{\bar{n}_{t}}(0)^2)}\right]
\ee
which is gaussian distributed  around $\epsilon=0$, i.e. $r=\bar r$, with variance ${\bar r(1-\bar r)+{w}'_{\bar{n}_{t}}(0)^2\over n_b}$.
Since
\be {w}'_{\bar{n}_{t}}(k)=i\sqrt{1\over 2\pi}\exp\left[
-{(i{k}+\bar n_t)^2\over 2}\right]
\ee
and $\bar r={w}_{\bar{n}_{t}}(0)$, $1-\bar r =1-{w}_{\bar{n}_{t}}(0) = {w}_{-\bar{n}_{t}}(0)$
we eventually get for the variance
\be (\Delta m)^2=n_b \, \phi(\bar{n}_{t})
\label{var} \ee
In the above we introduced the Kollines  function 
\be\phi(x) = {\cal {N}}(x) {\cal{N}}(-x) - {1\over {2\pi}}\exp[-x^2]
\ee
It is positive, very flat around $x=0$ (the first three derivatives vanish at $x=0$) and is essentially   zero when $|x|>3.5$ (see Fig. \ref{Kollines}).

\begin{figure}[htbp]
\epsfxsize=8cm
\centerline{\epsfbox{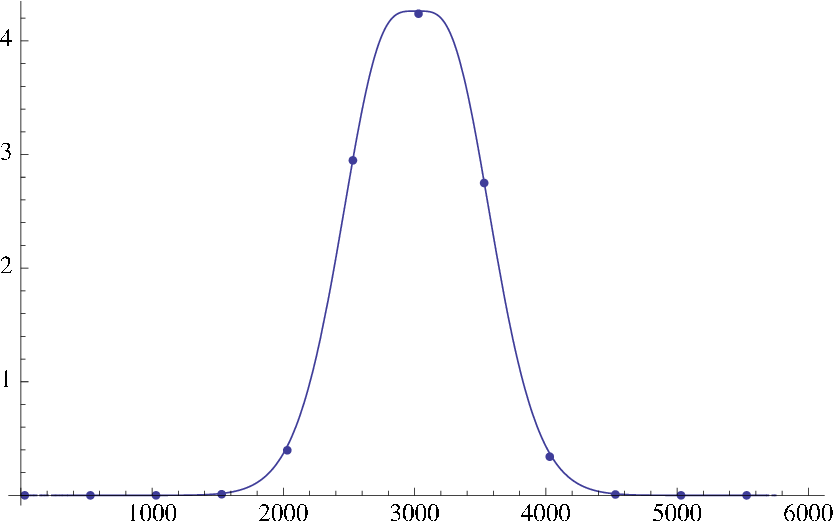}}
\caption{The standard deviation of the final rank of the virtual boat for $200$ boats racing  $30$ races : the continuous line is the square root of the Kollines function and the points are numerical simuations for a score $n_t$ ranging from $30$ to $6000$ by steps of $500$.}
\label{standardbis}
\end{figure}

It follows that when $|{n}_{t}-n_r(n_b+1)/2| \gg 3.5/\sqrt{\lambda}\,(\simeq 3.5 n_b\sqrt{n_r/12})$ the final rank has no fluctuation. It is only when   $|{n}_{t}-n_r(n_b+1)/2|<3.5/\sqrt{\lambda}$  that 
$\Delta m\simeq \sqrt{n_b}$ as illustrated in Fig. \ref{standardbis} for $200$ boats racing $30$ races.

\section{Small race number:  the case  $n_r=2$}\label{sec-2r}

The problem without the benefit of the large-$n_r$ limit becomes harder and, for generic $n_r$, is not amenable to an explicit solution. For
the case of few races, however, we can obtain exact results.

\begin{figure}[htbp]
\epsfxsize=8cm
\centerline{\epsfbox{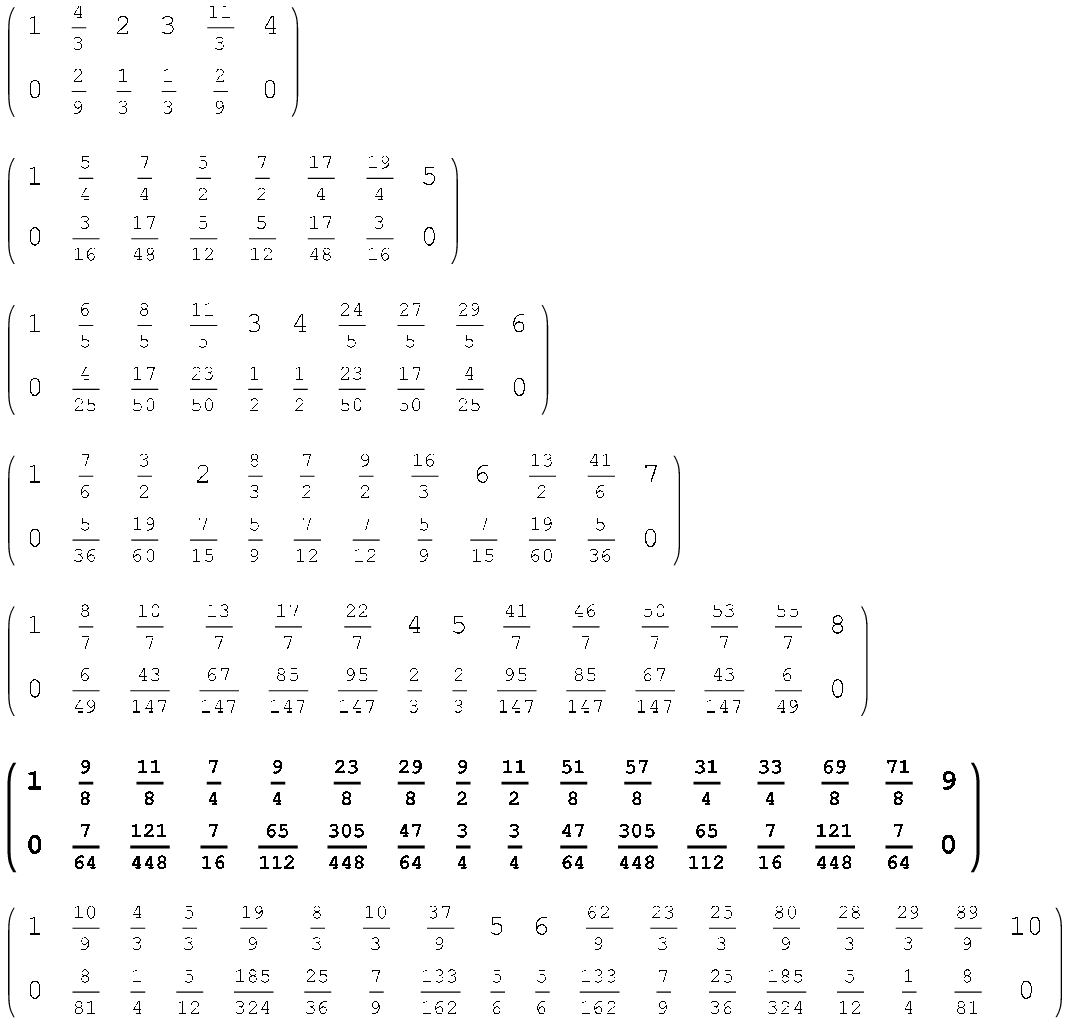}}
\caption{By complete enumeration of all permutations: the mean final rank and variance for $3, 4,..., 9$ boats and $2$ races. }
\label{2races}
\end{figure}

In the present section we deal with the case $n_r =2$, for which we can
find the exact solution. Fig. \ref{2races} displays the mean final ranks and  variances of the virtual boat for  $n_b=3, 4,.., 9$ boats racing 2 races. For a given $n_b$  the score of the virtual boat spans the interval $[2,2n_b+1]$. 

\subsection{Sketch and basic properties}

\begin{figure}
\begin{center}
\includegraphics[scale=.40,angle=0]{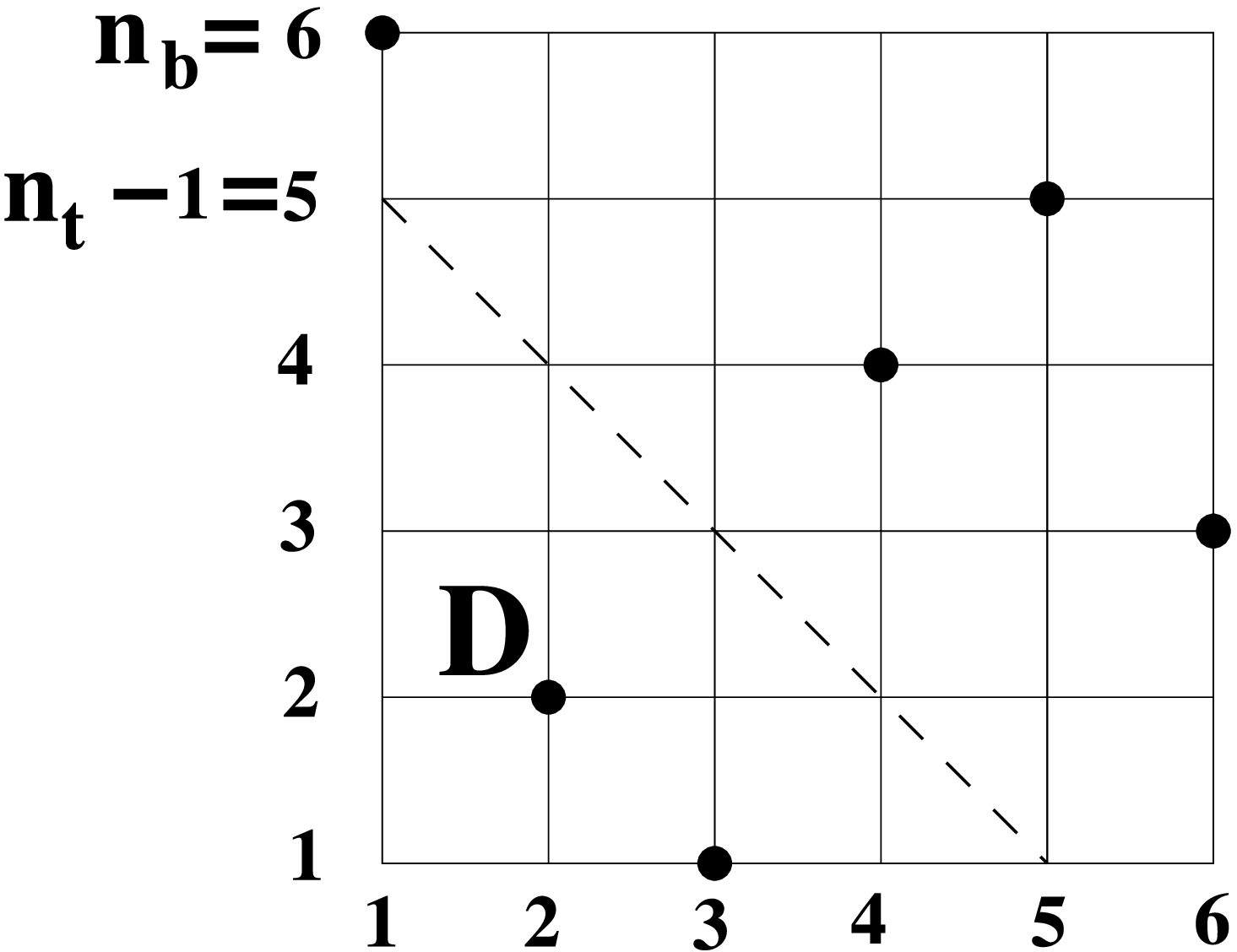}
\caption{The sketch of an event for $n_r=2$  and  $n_b=6$.  A boat is represented by a point whose coordinates are its ranks
   in the two races.  
 Here, we fix the score  $n_t=6$ of  the virtual boat (dashed
  diagonal). There are 2 sites occupied in  {\bf D}. Thus, the rank of the virtual boat is $m=3$ for this event.}\label{fig1}
\end{center}
\end{figure}

For two races, the situation can be sketched by using a $n_b \times
n_b$ square lattice as in Fig. \ref{fig1} for $n_b=6$.

The two coordinates correspond to the ranks of a  boat in each one
 of the two races. So, each boat will be represented by an occupied
 site. It follows that each line and each  column will be occupied
 once and only once. This leads to $n_b!$ possible configurations.

The  score  $n_t$ of the virtual boat is fixed and represented by the
dashed diagonal.  Let us call  $\bf D$ the domain under the diagonal. The rank of the virtual boat is equal to $m$ when $(m-1)$
sites are occupied in $\bf D$. We have obviously \ 
   $P_{n_t}(m)= \delta_{m,1}$ \ when $n_t \le 2$ and \
    $P_{n_t}(m)= \delta_{m,n_b+1}$  when $n_t \ge 2n_b+1$. Moreover,
    from symmetry considerations, 
\be\label{sym1}
P_{n_b+1-k}(m) = P_{n_b+2+k}( n_b+2- m) \; , \qquad k=0, 1, ...,n_b-1 
\ee
So, in the following, we will restrict $n_t$ to the range $2\le n_t \le
n_b+1$. In that case, it is easy to realize that only $(n_t-2)$
columns (or lines) are available in {\bf D}. This implies for $m$ the  restriction  $1\le m \le n_t-1$. 

We also observe that the distribution is symmetric for $n_t=n_b+1$ or
 $n_b+2$
\be\label{sym2}
P_{n_b+1}(m) = P_{n_b+1}( n_b+1- m)= P_{n_b+2}(m+1) \; , \qquad m=1, 2, ...,n_b 
\ee
\noindent
We will come back to this point later.

\subsection{Direct computations of  $P_{n_t}(m)$ for some  $m$}

\begin{figure}
\begin{center}
\includegraphics[scale=.40,angle=0]{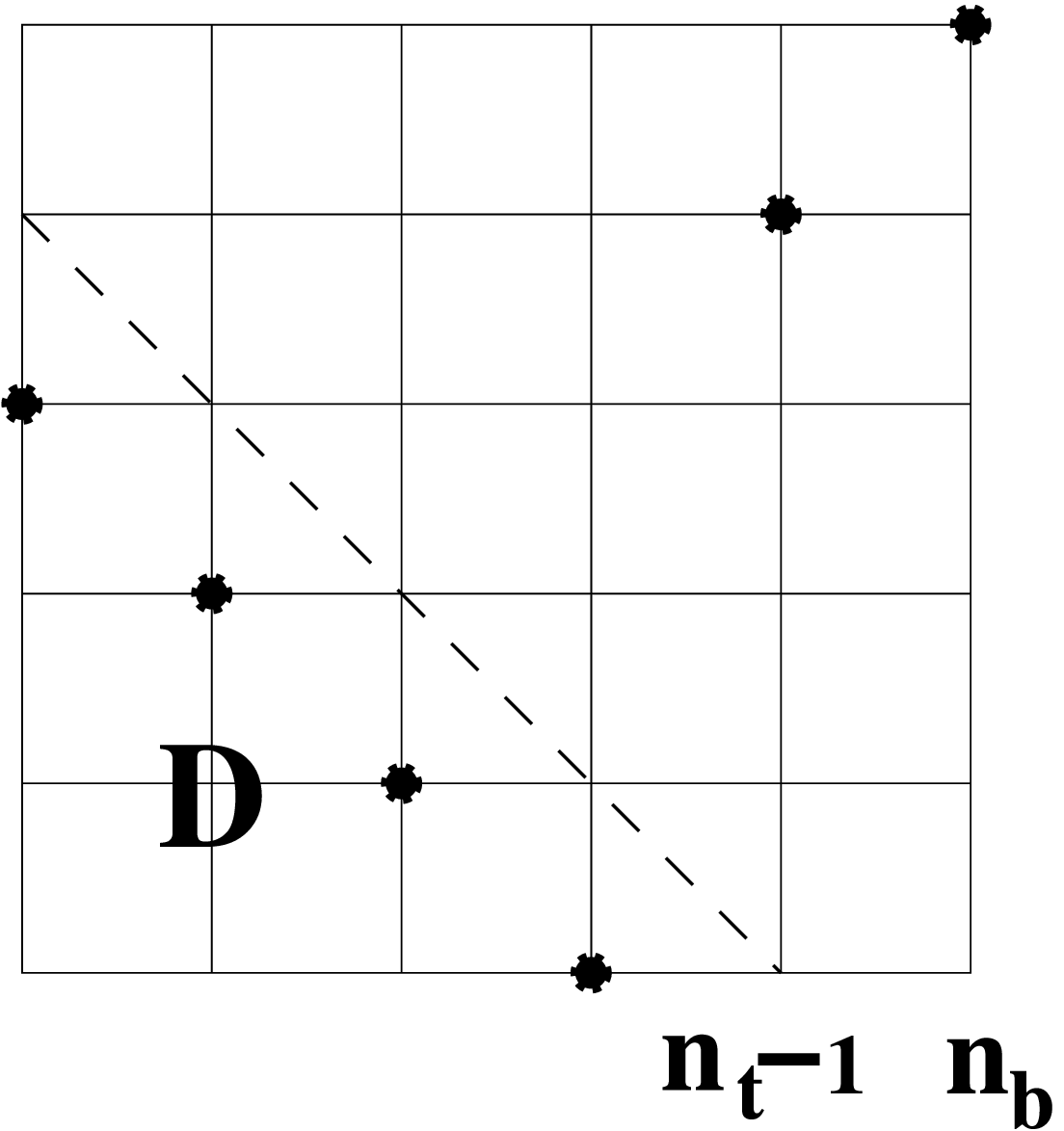}
\caption{A configuration contributing to $P_{n_t}(m=n_t-1)$. We have only one
  possibility for the $(n_t-2)$ occupied sites under the dashed diagonal.}\label{fig1b}
\end{center}
\end{figure}

For $m=n_t-1$, we observe (see Fig. \ref{fig1b}) that there is only one possibility to 
 occupy the $(n_t-2)$  sites in {\bf D}.

\noindent
The $(n_b-n_t+2)$ remaining
 occupied sites  are distributed randomly on the  sites of
 the  $(n_b-n_t+2)$  remaining lines and columns that are still available. 
  Denoting $(u \equiv n_b-n_t+2)$ one obtains
\be\label{nt1}
 P_{n_t}(m=n_t-1)= \frac{u!}{n_b!}  
\ee
\vskip.1cm

\begin{figure}
\begin{center}
\includegraphics[scale=.40,angle=0]{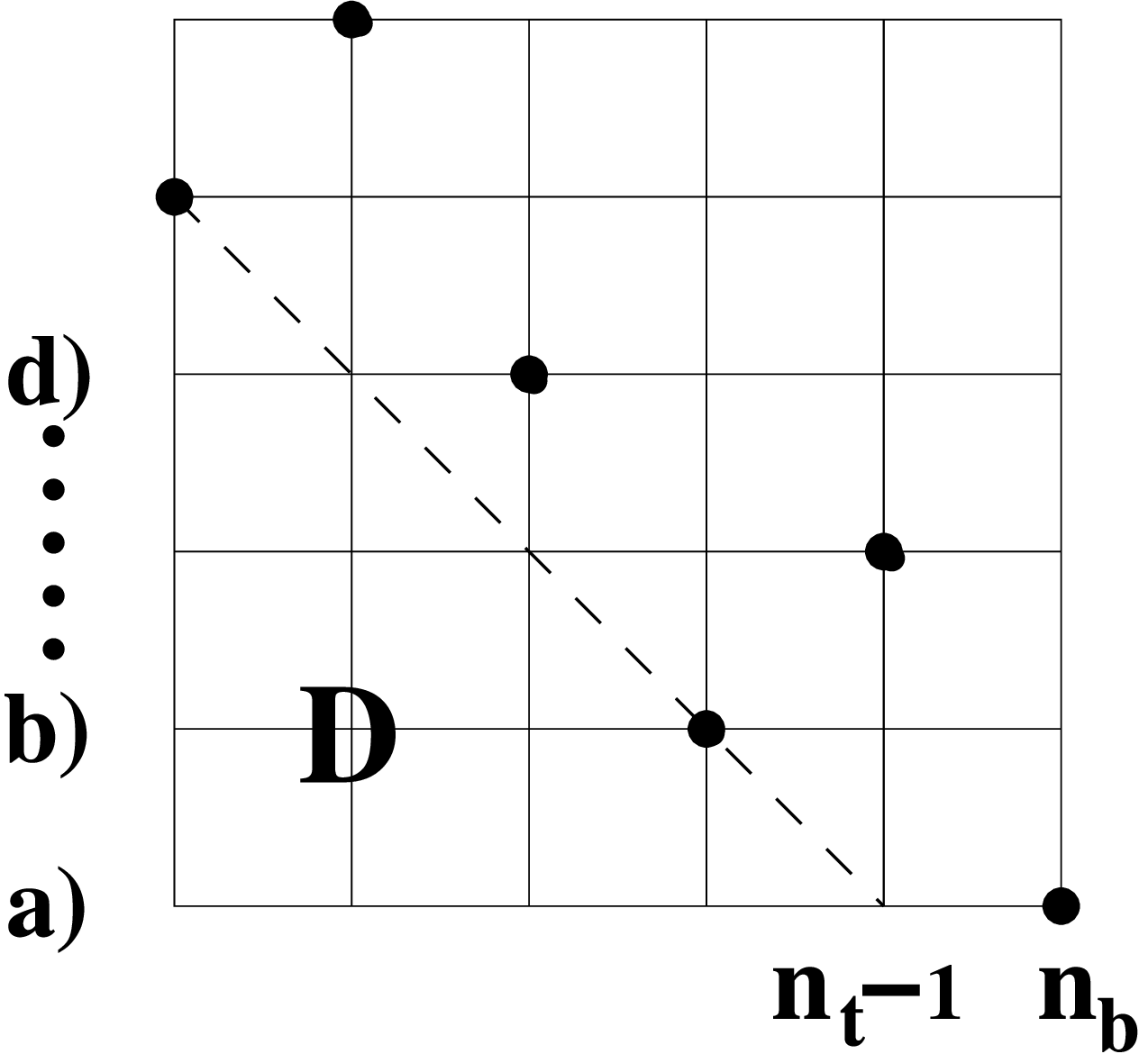}
\caption{A configuration contributing to $P_{n_t}(1)$. No occupied site
  belongs to {\bf D}. For each line a), b),
  ..., d), we have $n_b-n_t+2 $ possibilities for   the
 occupied sites. The remaining occupied  sites will generate the
 factor $P_{n_t}(n_t-1)$. For further explanations, see the text.}\label{fig2}
\end{center}
\end{figure}

Now, for $m=1$, there are no occupied sites in $D$. Let us fill
 (Fig. \ref{fig2}) the lines, starting from the bottom. On line (a),
  we have $n_b-n_t+2$ $(\equiv u)$ available sites; on line (b), we still
 have $u$ available sites (because of the site occupied in line
 (a)); and so on, up to line (d). Moreover, from the $u$ upper lines, we still get
a factor  $u!$.

\noindent
Finally
\be\label{m1}
 P_{n_t}(1)= P_{n_t}(n_t-1) \cdot \Phi_1(u) \qquad  \mbox{with}  \qquad \Phi_1(u) = u^{n_t-2}
\ee

\vskip.3cm

\noindent
It is easy to see, from the above considerations, that, for $1\le m
\le n_t -1$
\be\label{mg}  
 P_{n_t}(m)= P_{n_t}(n_t-1)\Phi_m(u)
\ee
where $\Phi_m(u)$  is a polynomial in $u$ with integer
values\footnote{This is not true for $ n_r \ge 3 $.}.

\vskip.3cm

For $m=2$, there is one occupied site, $B$, in {\bf D}. 

\begin{figure}
\begin{center}
\includegraphics[scale=.40,angle=0]{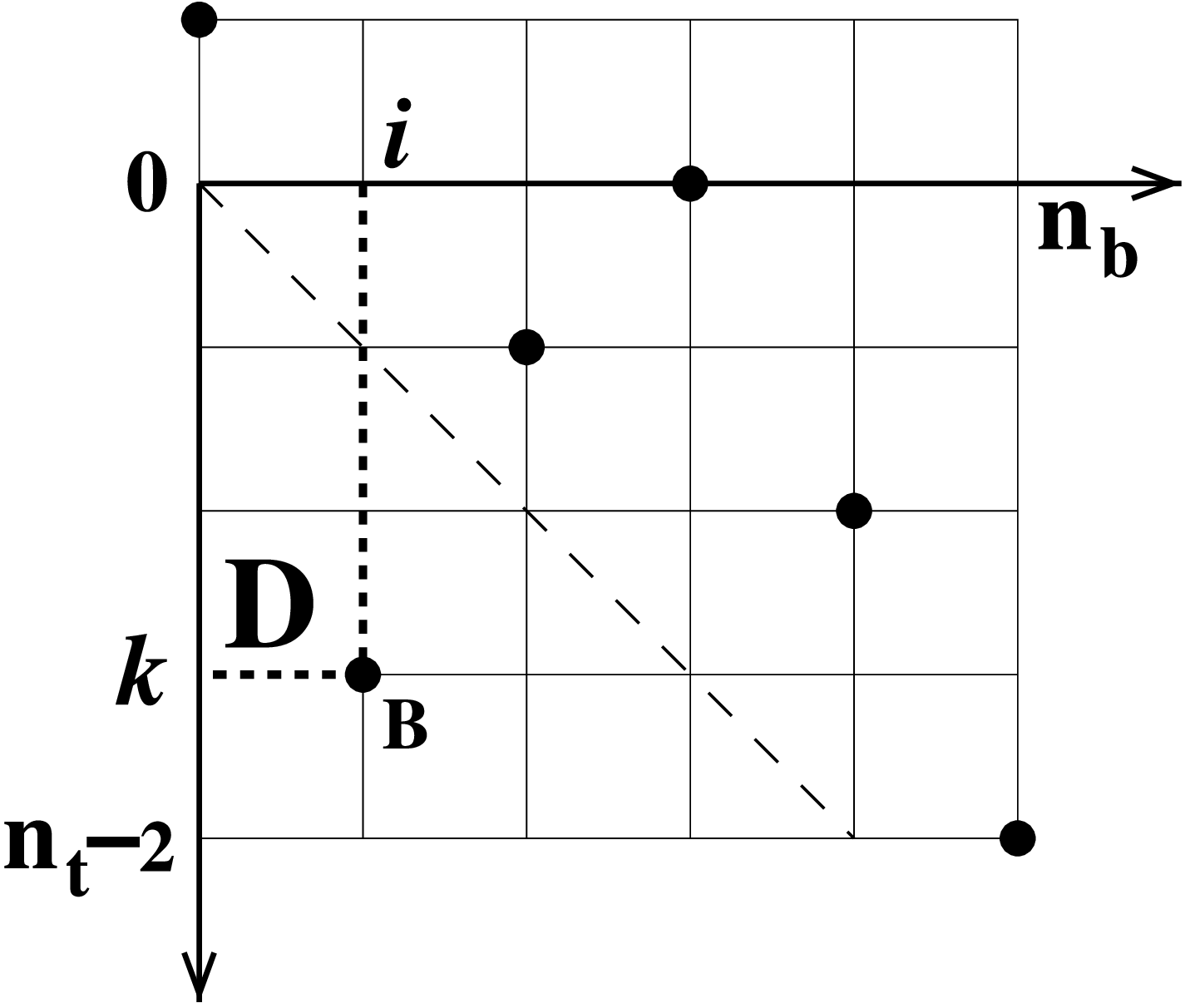}
\caption{ A configuration contributing to $P_{n_t}(2)$. The occupied site, B, in
  {\bf D}, has coordinates $i$ and $k$. For further explanations, see the text. }\label{fig3}
\end{center}
\end{figure}

\noindent
With the coordinates $(i,k)$ defined in Fig. \ref{fig3}, {\bf D} is
the domain $(0 \le i \le k-1; \; 1 \le  k \le n_t-2)$ so that
\be\label{m2}   
P_{n_t}(2)=P_{n_t}(n_t-1).\sum_{{\bf D}}u^{n_t-2-k}(u+1)^{k-i-1}u^i=P_{n_t}(n_t-1)\Phi_2(u)
\ee
with
\be\label{m22} 
\Phi_2(u)=(u+1)^{n_t-2}(u+1)-u^{n_t-2}(u+n_t-1)
\ee

\vskip.1cm

The computation for $m=3$ is more involved because the relative
position of the two occupied sites in {\bf D} plays an important
 role in the expression of the terms to be summed. One gets
\bea
\Phi_3(u) & = &  \frac{1}{2} \bigg[
(u+2)^{n_t-2}(u+1)(u+2)-2(u+1)^{n_t-2}(u+1)(u+n_t-1) \; +  \nonumber \\
 &&   + \;  u^{n_t-2}(u+n_t-1)(u+n_t-2) \label{m3}
\bigg]
\eea

\noindent
It is worth noting that, despite the apparent complexity of
$\Phi_3(u)$, the degree of $\Phi_m(u)$ decreases when $m$ increases.
  We will clarify this point later.

\vskip.2cm
 The case $m=4$ seems out of reach by direct computation and will not be
 pursued along these lines.

\subsection{Recursion relation and  solution of the case $n_r=2$}

Looking at  (\ref{m1}, \ref{m22}, \ref{m3}),
 we observe that, for $m\le  2$,
$\Phi_m(u)$ satisfies the recursion relation
\be\label{rec}
\Phi_{m+1}(u)= \frac{1}{m} \bigg( (u+1)\Phi_m(u+1)-(u+n_t-m)\Phi_m(u) \bigg)
\ee

\noindent
We will now show that (\ref{rec}) holds in general.

\begin{figure}
\begin{center}
\includegraphics[scale=.40,angle=0]{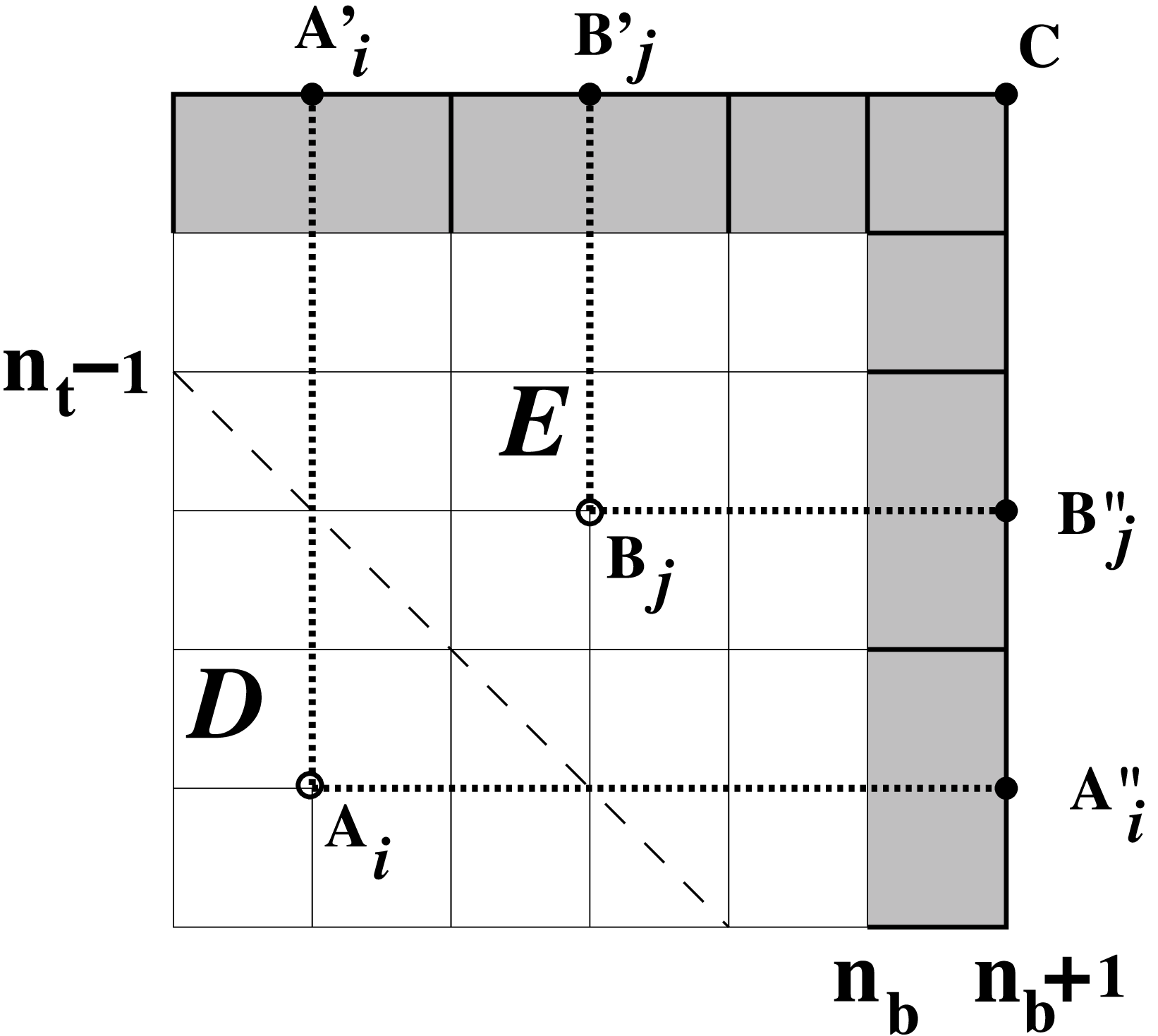}
\caption{ The 3 ways for producing a configuration contributing to
  $N'(m)$ (see the text for definition): i) start from a configuration contributing to
  $N(m+1)$, erase $A_i$ and add $A'_i$ and  $A''_i$; ii)  start from a configuration contributing to
  $N(m)$, erase $B_j$ and add $B'_j$ and  $B''_j$;  iii) start from a configuration contributing to
  $N(m)$ and add $C$. }\label{fig4}
\end{center}
\end{figure}

Let us write \  $ P_{n_t}(m)= \frac{u!}{n_b!} \Phi_m(u) =
\frac{N(m)}{n_b!}    $
 \ where $N(m)$ is the number of configurations of the $n_b \times
 n_b$ square  with $(m-1)$ occupied
 sites in {\bf D}.
Changing $n_b$ into $n_b+1$ (which amounts to changing $u$ into
$u+1$  while keeping $n_t$ unchanged), we call $P'_{n_t}(m)$ the new probability distribution
    $ P'_{n_t}(m)= \frac{(u+1)!}{(n_b+1)!} \Phi_m(u+1) =
\frac{N'(m)}{(n_b+1)!}    $ 
  \ where $N'(m)$ is defined like  $N(m)$ but for the $(n_b+1) \times
  (n_b+1) $ square lattice (Fig. \ref{fig4}).

\vskip.3cm

$N'(m)$ receives three kinds of contributions:

i) Let us consider a configuration contributing to $N(m+1)$ ($m$
occupied sites $A_i$ in {\bf D} -- see Fig. \ref{fig4}). The
replacement of  $A_i$ by   $A'_i$  and  $A''_i$ produces a
configuration contributing to $N'(m)$ (only $(m-1)$ occupied sites in
{\bf D}; all the columns and lines of
the biggest square are occupied once). Since we can choose any of the
$A_i$'s   before  applying this
 procedure, we get a contribution $m N(m+1)$ to $N'(m)$.

ii) Let us next consider a configuration contributing to $N(m)$ ($n_b +1-m$
occupied sites $B_j$ in {\bf E} -- see Fig. \ref{fig4}). By the same reasoning as in (i), we get $(n_b+1-m)N(m)$ configurations for $N'(m)$.

iii) To each configuration  contributing to $N(m)$, we can add an
occupied site in $C$ (see Fig. \ref{fig4}). This produces the
contribution $N(m)$ to $N'(m)$.

\noindent
Summing the above contributions leads to
\be\label{np}
N'(m) = mN(m+1) + (n_b+2-m)N(m)
\ee
Reverting back to $\Phi_m $'s, it is straightforward to get (\ref{rec}).
Equations (\ref{m1}) and (\ref{rec}) prove that  $\Phi_m(u) $ has degree $n_t-m-1$.

  Finally, solving the recursion equation, we get the exact
  solution for $n_r=2$
\be\label{sol}
P_{n_t}(m)=(n_b+1)\sum_{k=0}^{m-1}(-1)^k(n_b-n_t+m-k+1)^{n_t-2}\frac{(n_b-n_t+m-k+1)!}{
 k! (n_b-k+1)! (m-k-1)!    }         
\ee
with \ $2 \le m+1   \le n_t \le n_b + 1 $ understood.
We have checked (\ref{sol}) by a complete enumeration of the
permutations up to $n_b$ and  $n_t=10$.
 
Let us discuss the case $n_t=n_b+1$. Equation (\ref{sol}) narrows down to 
\be\label{solsym}
P_{n_t}(m)=n_t\sum_{k=0}^{m-1}  (-1)^k  \frac{(m-k)^{n_t-1}}{ k! (n_t-k)! }  
\ee
The moments are
\bea
\langle m^n\rangle &=& n_t \left(\frac{\partial}{\partial \lambda'   }
\right)^n\bigg\vert_{\lambda' =0}
\left(\frac{\partial}{\partial \lambda   }
\right)^{n_t-1}\bigg\vert_{\lambda =0}\sum_{m=1}^{n_t-1}\sum_{k=0}^{m-1} \frac{(-1)^k}{ k! (n_t-k)! } 
 e^{\lambda (m-k)   } e^{\lambda' m  } \nonumber \\
 &=& \frac{1}{(n_t-1)!}\left(\frac{\partial}{\partial \lambda'   }
\right)^n\bigg\vert_{\lambda' =0}
\left(\frac{\partial}{\partial \lambda   }
\right)^{n_t-1}\bigg\vert_{\lambda =0}  
\left[
\frac{ (1-e^{\lambda'})^{n_t}-( e^{\lambda +  \lambda'} - e^{\lambda'}
  )^{n_t}      }
{  1-e^{\lambda +    \lambda'} }
\right]\label{momsym}
\eea
and in particular
\bea
\langle m\rangle &=& \frac{n_t}{2}    \label{mom1} \\
\langle (m - \langle m\rangle   )^2\rangle  &=& \frac{n_t}{12} \label{mom2} \\
\langle (m - \langle m\rangle   )^3\rangle  &=& 0 \label{mom3}
\eea
We recover the fact that $P_{n_t}(m)$ is symmetric. These results will be
  especially  useful  in the next section.

\subsection{Computations of the first three moments for $n_t \le n_b$ }

Starting from the equation (\ref{np}), we get
\be\label{recp}
mP_{n_t}(m+1)=(n_b+1)P'_{n_t}(m)-(n_b+2-m)P_{n_t}(m)
\ee
(recall that $P'_{n_t}(m)$ is the same as $P_{n_t}(m)$  but for $n_b$  changed
into $n_b+1$). Multipying both sides of (\ref{recp}) 
  by  $m^k$  and summing over $m$,   the recursion equation for
  the moments follows
\be\label{recm}
(n_b+1-k)<m^k>+\sum_{p=0}^{k-1} \frac{ (-1)^{k+1-p}  (k+1)!  }{p!(k+1-p)!}<m^p>=(n_b+1)<m^k>'
\ee
($<...>$ refers to  $n_b$ and   $<...>'$  to  $n_b+1$    ).

\vskip.1cm

\noindent
For $k=1$, setting $Z_{n_b}=n_b<m>$, we get $Z_{n_b}-Z_{n_b-1}=1$ and,
finally, $Z_{n_b}=Z_{n_t-1}+n_b-n_t+1$. Computing $Z_{n_t-1}$ with (\ref{mom1}), we
obtain the first moment
\be\label{mom11}
<m> = 1+ \frac{(n_t-1)(n_t-2)}{2n_b}
\ee
The other moments are obtained in a similar way. Equations (\ref{mom2}),
(\ref{mom3}) and  (\ref{recm})  lead to:
\bea
<(m - <m>   )^2>  &=&  \frac{(n_t-1)(n_t-2)}{12n_b^2(n_b-1)} 
 \left[  3 n_t^2-n_t(9+8n_b)  +   6(n_b+1)^2
 \right], \quad  n_b\ge 2  \label{mom22}    \\
<(m - <m>   )^3> &=&    \frac{(n_t-1)(n_t-2)(n_t-n_b-1)^2(n_t-n_b-2)^2}
{2n_b^3(n_b-1)(n_b-2)}   \; , \qquad  n_b\ge 3    \label{mom33}
\eea
As expected,    $<(m - <m>   )^3>$ vanishes for $n_t=n_b+1$ or  $n_b+2$
(the distribution is symmetric);  $<(m - <m>   )^2>$  and $<(m - <m>
)^3>$  vanish for $n_t=1$ or $2$ ($P_{1,2}(m) = \delta_{m,1}     $).

\section{The case  $n_r\ge 3$}

\begin{figure}[htbp]
\epsfxsize=8cm
\centerline{\epsfbox{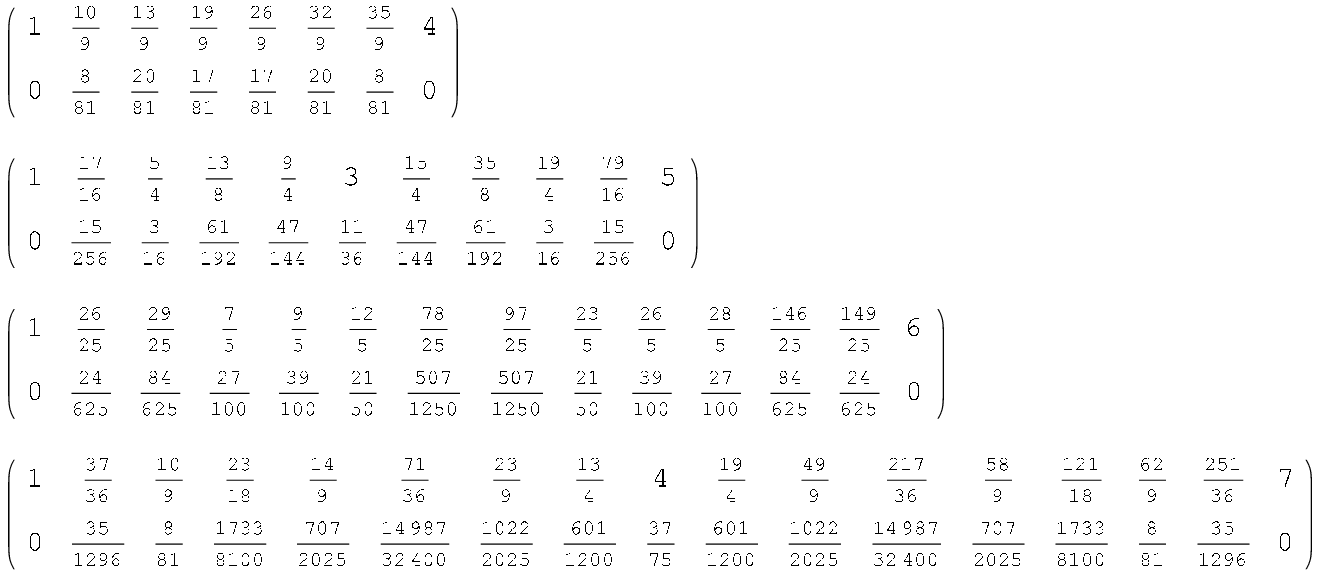}}
\caption{By complete enumeration of all permutations: the  mean final rank and variance for 3, 4, 5 and 6 boats and 3 races. }
\label{3race}
\end{figure}

For the case of three or more races the problem is more complex. We
can, however, establish some partial exact results. Fig. \ref{3race}  demonstrates the stituation for three races, displaying the mean final ranks and variances of the virtual boat for  $n_b=3,4,5,6$ boats. 
The score of the virtual boat spans the interval $[3,3n_b+1]$. 

For $n_r =3$ and $n_t\le n_b+2$, we established and checked
numerically the recursion relation
\be
N'(m)=(m+1)mN(m+2)+m(2n_b-2m+3)N(m+1)+(n_b-m+2)^2N(m) 
\ee

More generally, for $n_r \ge 3$, we obtained the expression
\be
<m>= 1+  \frac{(n_t-1)!}{n_b^{n_r-1} (n_t-1-n_r)! n_r!  } \qquad
\mbox{for} \qquad n_r \le n_t \le n_b+n_r-1
\ee
 
\section{Two races with the worst individual rank dropped}

We conclude our analysis with a variant of the original problem,
also used in competitions, for the specific case of two races.

Specifically, suppose that, for each boat, we drop the greatest rank (worst result) obtained in the two races. For instance, if the boat $i$ had ranks
   $n_{i,1}=2$ and  $n_{i,2}=5$, we only retain the score  $n_i=2$. The
   virtual boat has a fixed score $n_t$ in the range $[1,n_b+1]$  and, as before, 
its rank is $m$ when $(m-1)$ boats have scores $n_i$ smaller than $n_t$.   

It is obvious that $m\ge n_t$.
Indeed, without loss of
      generality, we can consider that the ranks $n_{i,1}$ obtained in
      the first race are arranged in natural order: $ \{1,2, ...,
      n_b-1, n_b\} $, ie $n_{i,1}=i$. (We will keep this order all
      along this section). 
   Now, from $ n_i \le n_{i,1} $,
      it is easy to realize that, at least $(n_t-1)$ boats will have
      scores $n_i$ smaller than $n_t$, thus $ m \ge n_t $.

\noindent
Defining the ordered sets \ $A= \{1, 2, ..., n_t-2, n_t-1 \} $   and 
   \ $B= \{n_t, n_t+1, ..., n_b-1, n_b \} $, we see that, taking, for
   the ordered\footnote{Here, ``ordered" does not mean ``in natural
     order" but simply that we take into account the order when we
     enumerate the elements of the set (i.e., \,   $\{a, b, ... \} \ne \{b, a,
     ... \} $).
 } set of ranks $r_{i,2}$ in the second race, any permutation of $A$  
   (for instance  $ \{n_t-2, 2,    1, ...,  n_t-1 \} $ ) followed by
   any permutation of $B$  
   (for instance  $ \{n_b-1, n_t,    n_t+1, ...,  n_b \} $ ), we 
      construct all the configurations leading to $m=n_t$. The number
      of such configurations is $(n_t-1)! \times (n_b-n_t+1)!$. Dividing by
      the total number of configurations $n_b!$, we get:
\be\label{pntnt}
   P_{n_t}(n_t)=\frac{1}{ \left( \begin{array}{c} 
  n_b \\
     n_t -1  \end{array}    \right)  } 
\ee

For  $m > n_t$, we start from the naturally ordered sets $A$ and $B$ and
 exchange  $(m-n_t)$ elements of $A$ with  $(m-n_t)$ elements of $B$ 
 (of course, $m-n_t \le n_t - 1$  and  $m-n_t \le n_b - n_t + 1$). So,
 we get the sets $A'$ and $B'$. Taking, for the ordered  set of ranks in the
 second race, any permutation of $A'$   followed by
   any permutation of $B'$, we get all the configurations leading to
   the rank $m$ for the virtual boat. We eventually obtain a
   hypergeometric law for the random variable $(m-n_t)$
\be\label{pntm}
   P_{n_t}(m)=\frac{
\left( \begin{array}{c} 
  n_t -1      \\
    m - n_t \end{array}    \right)
\left( \begin{array}{c} 
  n_b - n_t +1 \\
  m-   n_t   \end{array}    \right)
}
{ \left( \begin{array}{c} 
  n_b \\
     n_t -1  \end{array}    \right)  } 
\ee
with \  $n_t \le m \le \mbox{min} \{2n_t - 1, \; n_b+1    \} $  

Of course, this probability density is quite different from the one
obtained in (\ref{sol}). In particular, it is interesting to note
that the distribution (\ref{pntm}) is unchanged when we replace, 
 simultaneously,  \
 $n_t$ by $n_t'=n_b+2-n_t$ and \  $m$ by $m'=m+n_b+2-2n_t$
\be\label{newrule}
  P_{n_t'}(m') =  P_{n_t}(m)
\ee
(Note  that $ n_t'-1 = n_b + 1 - n_t $, $ n_b - n_t'+ 1 =
n_t - 1 $ and $ m' - n_t' = m - n_t $. So, from (\ref{pntm}), $P_{n_t}(m)$ is unchanged.)

When $n_b$ is even, the distribution is symmetric for
$     n_t=  \frac{n_b}{2}  +  1$. Indeed

\be\label{paire}
P_{\frac{n_b}{2}+1}(m)=\frac{
\left( \begin{array}{c} 
  \frac{n_b}{2}     \\
    m - \frac{n_b}{2}  -  1  \end{array}    \right)^2
}
{ \left( \begin{array}{c} 
   n_b \\
   \frac{n_b}{2}    \end{array}    \right)  } = P_{\frac{n_b}{2} +  1}
  \left(\frac{3 n_b}{2}  + 2
- m  \right) \; , \qquad    \frac{n_b}{2} + 1  \le m \le n_b+1 
\ee

 The  moments of  (\ref{pntm}) are
\bea
<m> &=&   n_t + \frac{(n_t-1)(n_b-n_t+1)}{n_b} \label{mom11n} \\
<(m - <m>   )^2>  &=&  \frac{(n_t-1)^2(n_b-n_t+1)^2}{n_b^2(n_b-1)} 
   \; , \qquad  n_b\ge 2  \label{mom22n}    \\
<(m - <m>   )^3> &=&  \; - \;  \frac{ (n_t-1)^2(n_b-n_t+1)^2 (n_b-2n_t+2)^2       }
{n_b^3(n_b-1)(n_b-2)}   \; , \qquad  n_b\ge 3    \label{mom33n}
\eea
consistent with (\ref{newrule}).
  Moreover, as expected, $<(m - <m>   )^2>$  and  $   <(m - <m>   )^3> $ 
  vanish for $n_t=1$ and  $n_t=n_b+1$. Finally,  $   <(m - <m>   )^3> $ vanishes
  for  $   n_t=  \frac{n_b}{2}  +  1$   when $n_b$ is even
  (the distribution is symmetric, see   (\ref{paire})).

\section{Conclusions}

We demonstrated that the problem of determining the final rank
distribution for a boat in a set of  races given its total score can be explicitly solved in two distinct situations:
for a large number of races, and for a few (2 or 3) races. We also
demonstrated that the ``Statistical Curse of the  Second Half Rank" effect can be attributed to statistical averaging in the case
of many races.

Although we obtained our results in the context and language of boat
racing, they are clearly applicable in several similar situations,
such as, e.g., student ranks based on their results in many exams
or quizes, rank of candidates for positions or awards when they
are reviewed and ranked by many independent evaluators, and voting
results when voters submit a rank of the choices. 

There are many open issues and unsolved problems for further
investigation. The exact
result for an arbitrary number of races (greater than 2) is not
known. Further, the obtained results are based on the simplifying assumption
that all boats are equally worthy (all ranks in each race are
equally probable). One could examine the situation in which boats
have {\it a priori} different inherent worths, handicapping the
probabilities for the ranks, and see to what extent the
``statistical curse" effect also emerges. Finally, the relevance and
relation of our results with well-known difficulties in rank
situations, such as Arrow's Impossibility theorem \cite{arrow,g}, would be an interesting topic for further investigation.

\vspace{1cm}

{\bf Acknowledgements:} S.O. acknowledges useful conversations with H.Hilhorst, M. Leborgne, S. Mashkevich and S. Matveenko, and thanks the City College of New York for its hospitality during the completion of this work. A.P. acknowledges the hospitality of CNRS and Universit\'e Paris-Sud during the initial stages of this work. The research of A.P. is supported by an NSF grant.

\end{document}